# Implementation of Connectivity and Handover through Wireless Sensor Node based Techniques

[1]N.S.V.Shet, [2]Prof. K.Chandrasekaran, [3]Prof. K.C.Shet
shekar_shet@yahoo.com
[1,2,3]NITK,Surathkal

**Abstract**

**In this paper a scheme for handoff and connectivity, based on wireless sensor nodetechniques is proposed. Scenes are created in Qualnet and simulated for a simple case. Results are discussed.**

## 1. INTRODUCTION

A wireless sensor network (WSN) is a wireless network consisting of spatially distributed autonomous devices using sensors to co-operatively monitor physical or environmental conditions, such as temperature, sound, vibration, pressure, motion or pollutants, at different locations. The development of wireless sensor networks was originally motivated by military applications such as battlefield surveillance. However, wireless sensor networks are now used in many industrial and civilian application areas, including industrial process monitoring and control, machine health monitoring, environment and habitat monitoring, healthcare applications, home automation, and traffic control.

In addition to one or more sensors, each node in a sensor network is typically equipped with a radio transceiver or other wireless communications device, a small microcontroller, and an energy source, usually a battery. The envisaged size of a single sensor node can vary from shoebox-sized nodes down to devices the size of grain of dust, although functioning 'motes' of genuine microscopic dimensions have yet to be created. The cost of sensor nodes is similarly variable, depending on the size of the sensor network and the complexity required of individual sensor nodes. Size and cost constraints on sensor nodes result in corresponding constraints on resources such as energy, memory, computational speed and bandwidth. A sensor network normally constitutes a wireless ad-hoc network, meaning that each sensor supports a multi-hop routing algorithm (several nodes may forward data packets to the base station). In computer science and telecommunications, wireless sensor
Networks are an active research area. [1]

## 2. WSN TECHNIQUES FOR HANDOFF

In this research we propose a scheme titled "WSN based handoff techniques for seamless roaming in wireless communication". The flow diagram is shown in figure 4. Here each 'mote' (motes are deployed randomly in remote areas like forests, areas with less habitation etc.) is equipped with the capability to detect and recognize a mobile station. Whenever a mobile station (MS) is unable to communicate to the Base station (BS) due to the non availability of signals it searches for nearby motes. Motes in turn, acknowledge the mobile nodes about its presence. These motes are already having an idea about the existence/location of other motes in its vicinity. These motes forward the MS request to other motes until the request reaches the nearest BS. BS can now steer their narrow beam antennas to the exact location of these MS with enhanced signal strength. In cases where the steering of antennas is not feasible satellite based routing is tried. In the present days antenna steering are not mechanical movement, but steering is achieved with the use of array of antennas. Once the link is established all the motes involved in the routing are out of the scenario, to conserve power.

## 3. HANDOFF IMPLEMENTATION USING WSN TECHNIQUES

The flow diagram shows the procedure involved in implementing handoff while a Base station is not available for immediate access. The MS searches for a near by WSN mote and sends a message that it needs to find a base station. The mote in turn forwards this message to its neighbor. The mote knows the location of its neighbor.

Motes search for BS by forwarding the message to all possible neighbors along the route until a BS is located. The message contains the exact location of MS. The BS on getting the information that a MS needs handoff, contacts the area MSC. MSC makes a decision as to which is the best BS that can communicate with this MS.

## 4. SCENARIOS TO DEMONSTRATE WSN HANDOFF TECHNIQUE

The BS starts a process of steering its antenna towards the MS. If steering and communicating is not possible, then the information is sent to satellite stations to locate the MS and start communication.

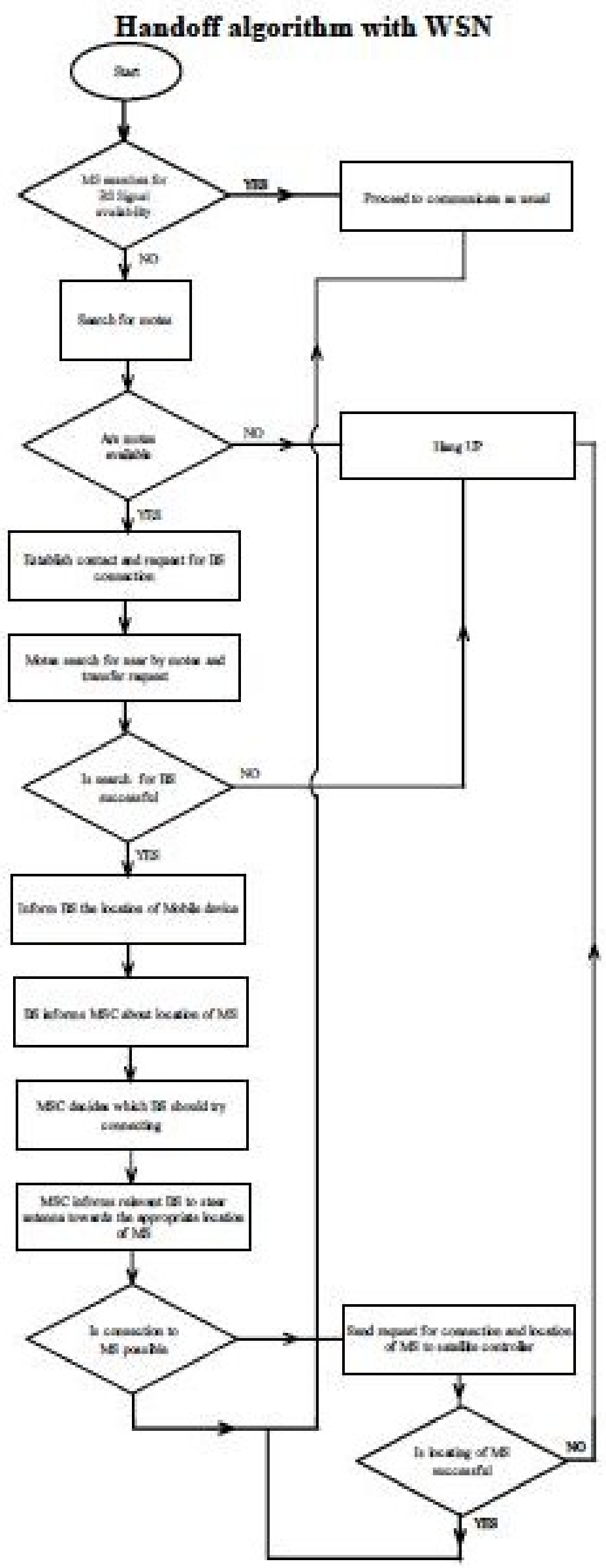

Figure 1 – Handoff algorithm with WSN

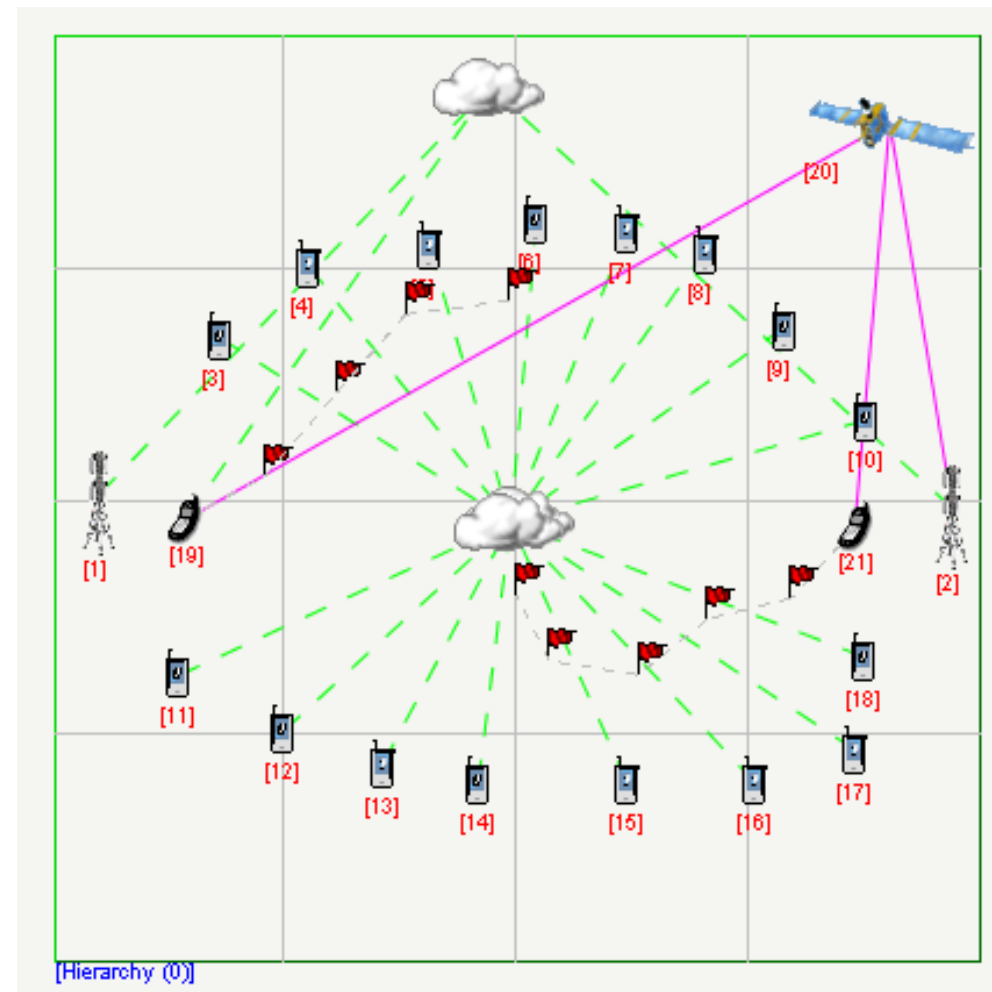

Figure 2

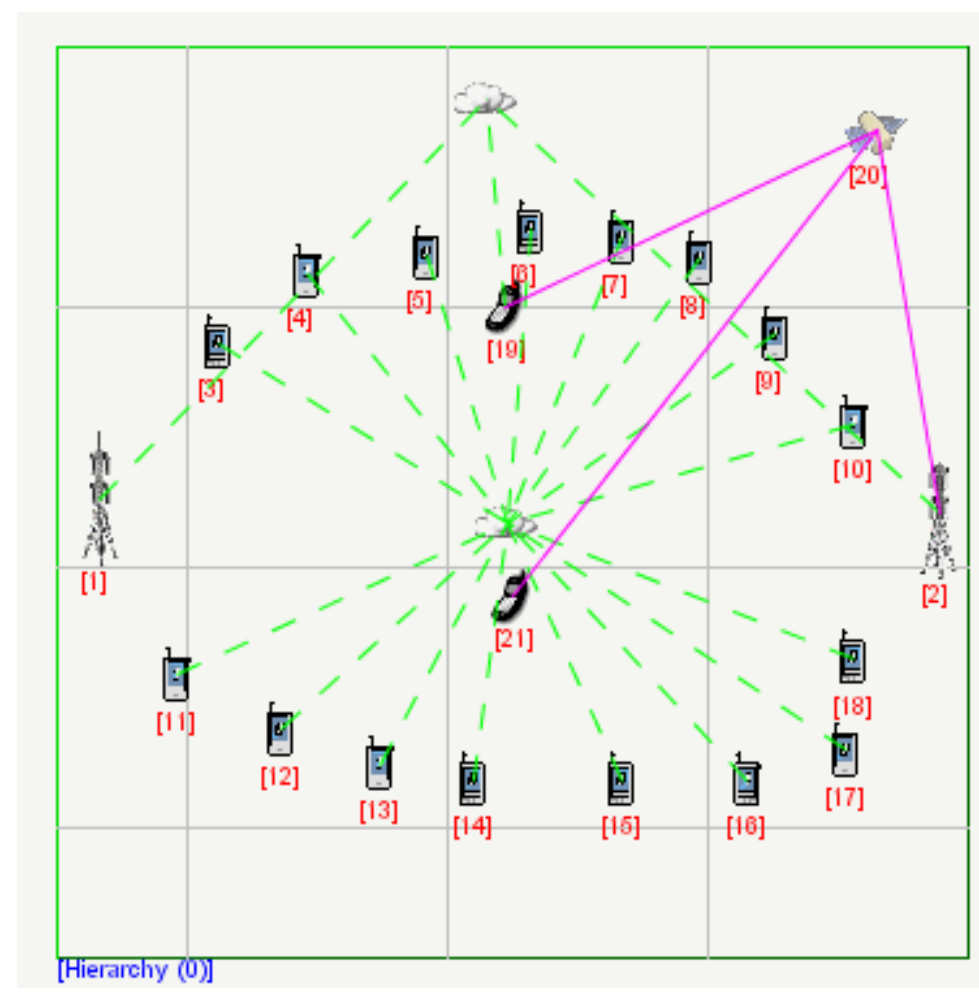

Figure 3

**Scenarios to demonstrate WSN handoff Technique**

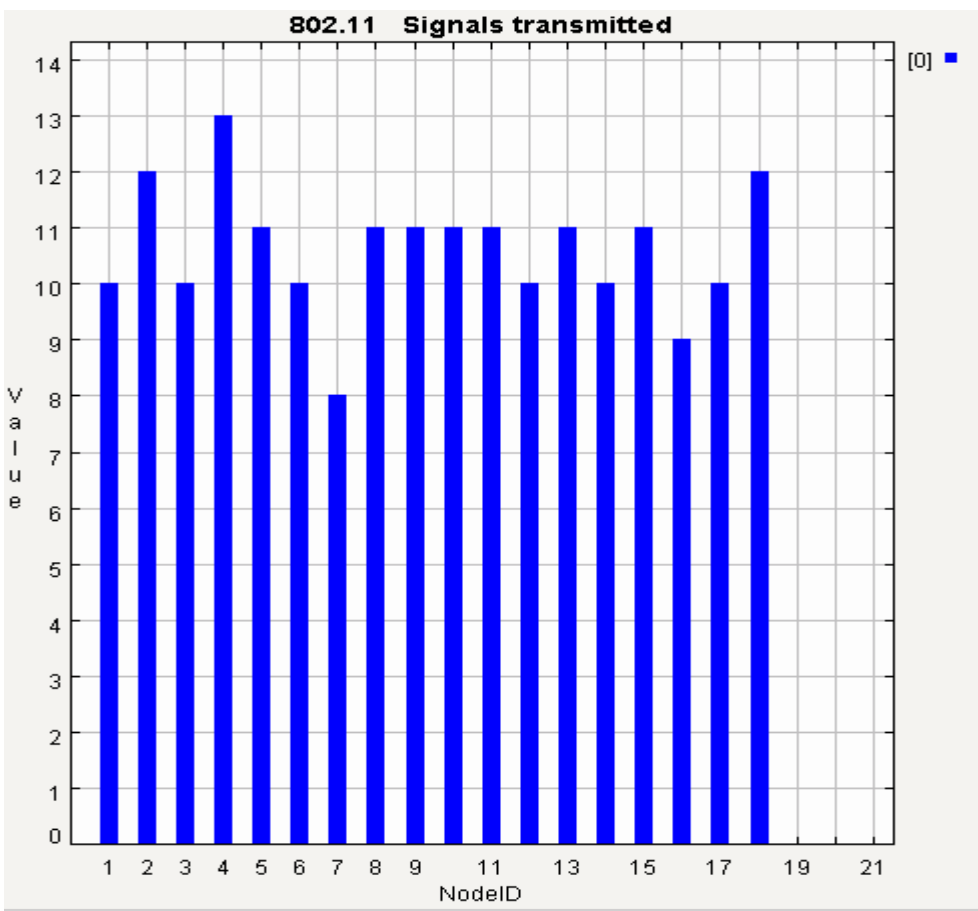

Figure 4

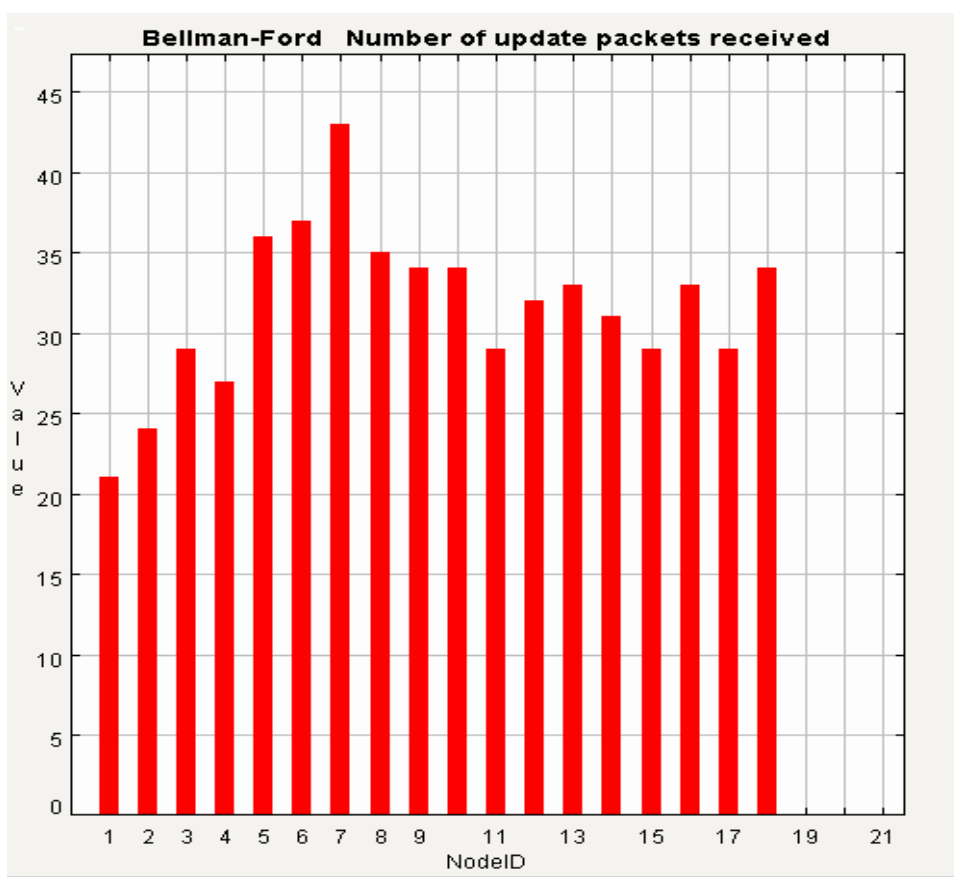

Figure 5

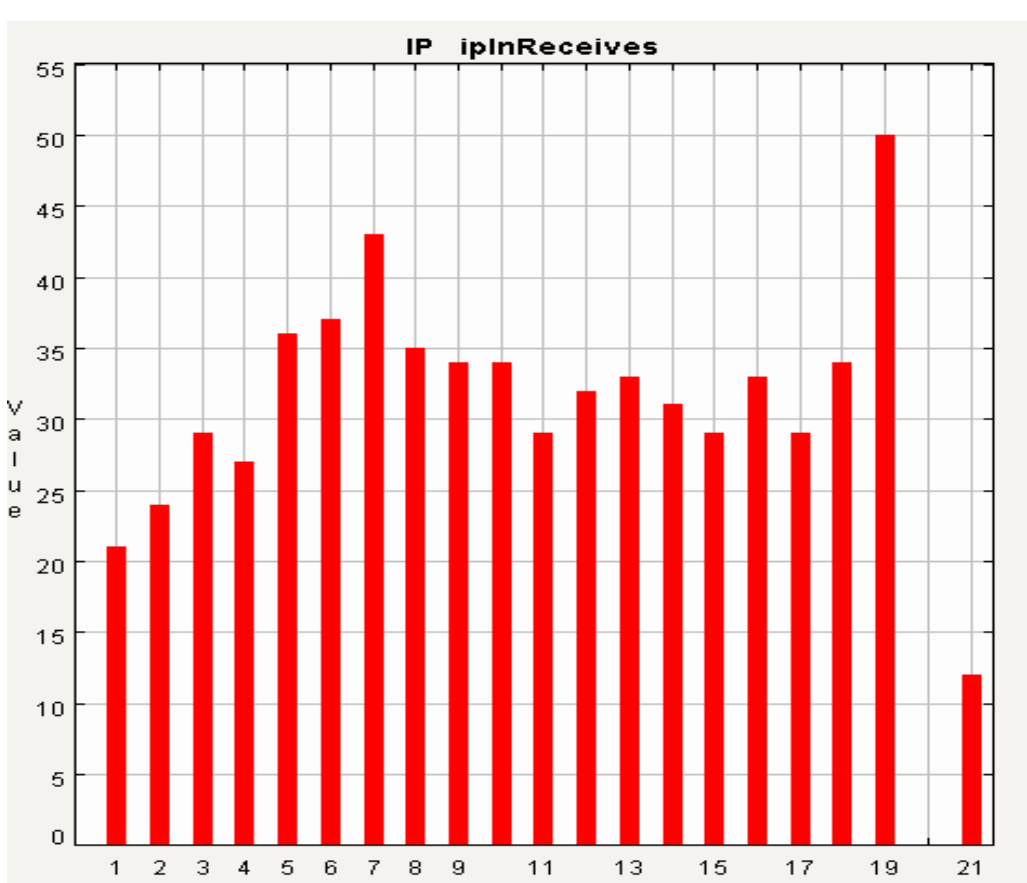

Figure 6

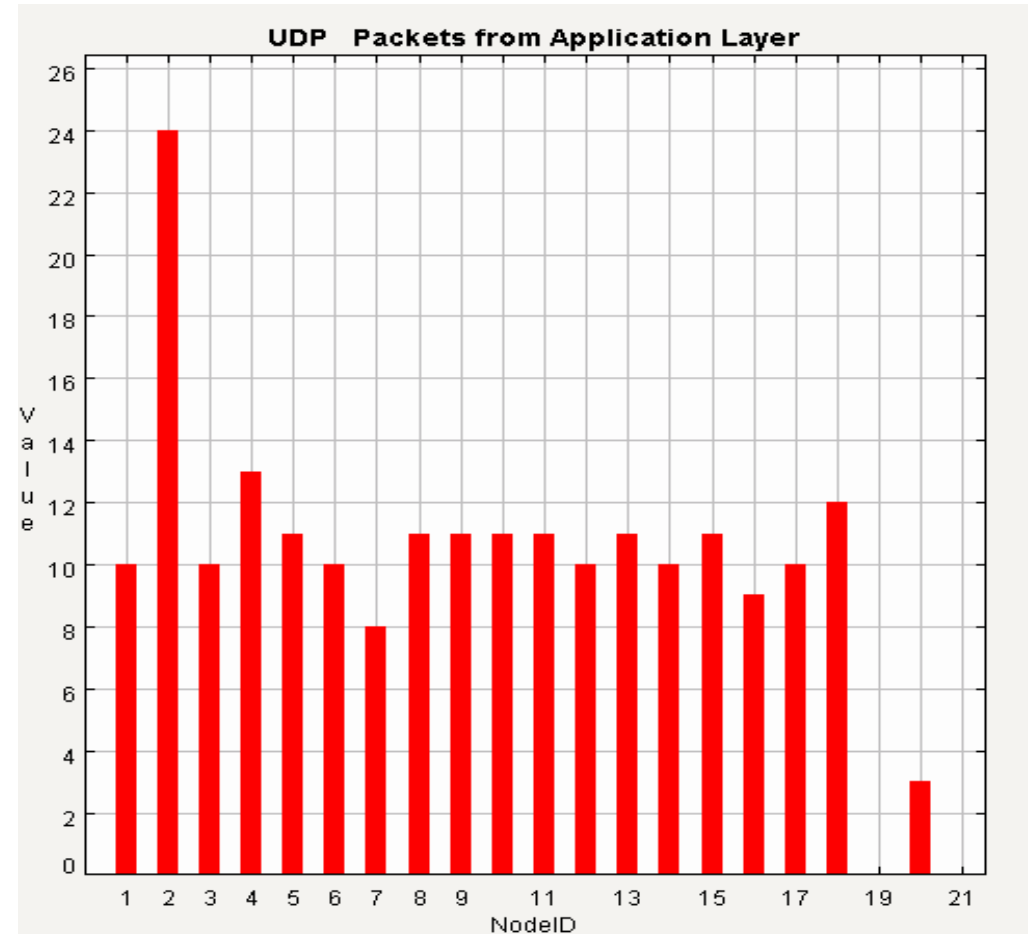

Figure 7

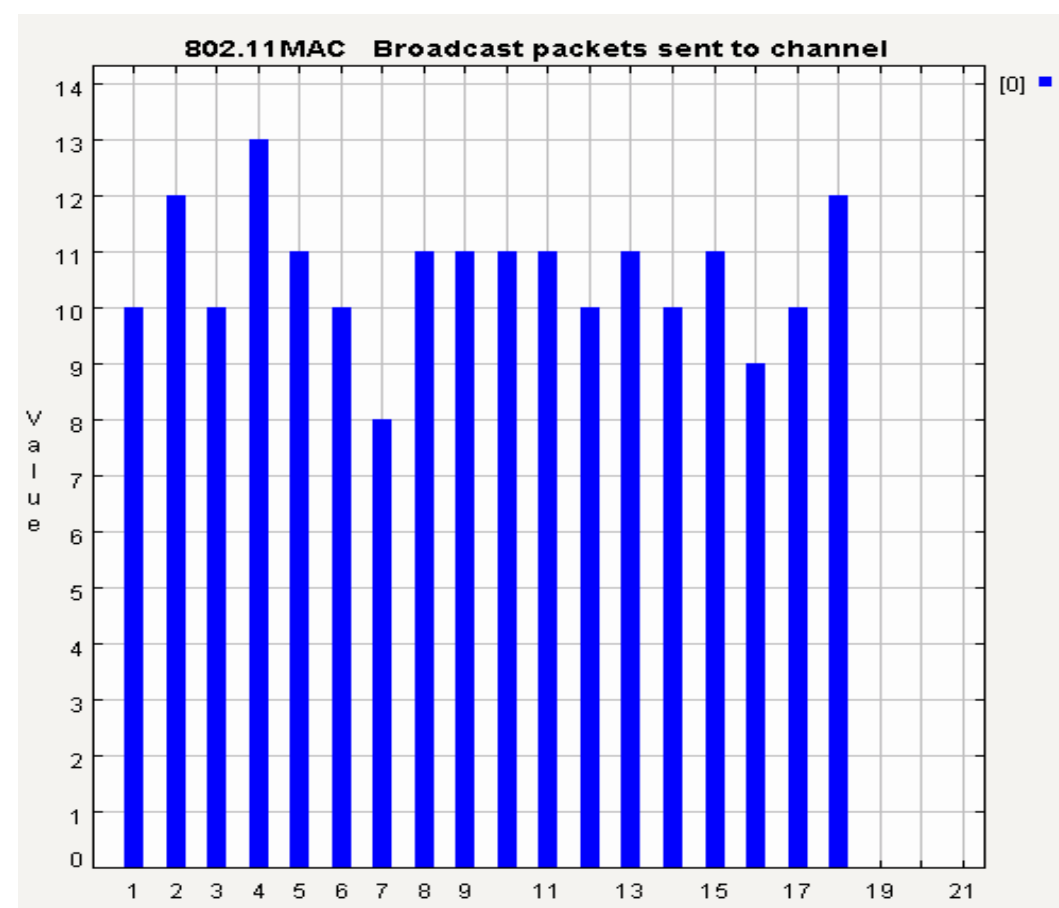

Figure 8

Figure one shows a scene that is created with two BS's, two MS's, 16 WSN nodes, one satellite station and way points for MS movement path. The MS's move such that they reach half way on the route towards the opposite BS and halt there as in fig. 2. At this point the mobile should be detected by the satellite station.

The results show the outputs at the various OSI layers. Figure 3 shows the signals transmitted. Fig 4 highlights the broadcast packets sent over the channel. IP packets received are clear in fig. 5. User defined packets from application layer are at fig 6. Number of updated packets received as per the Bellman ford algorithm is in Fig. 7.

The changes in various parameters in the QSI layers with wireless sensor nodes and without wireless sensor nodes are listed below.

**Physical Layer (802.11)**

- Signal transmitted with WSN is more.
- Signal received and forwarded to MAC increased from 6 units to 7 units.
- Signal locked on by Physical layer increased from 6 units to 13 units.
- Signal received with errors increased from nil to 6 units – undesirable.

**MAC Layer (802.11 MAC)**

- Packets from network layer increased.
- Broadcast packets sent to channel increased.
- Broadcast packets received clearly are increased.

**MAC Layer 802.11 DCF (Distributed Co-ordinate function)**

- Broadcast signals sent increased.
- Broadcast signals received increased.

**MAC Layer Link**

- No. of frames sent increased.
- No. of frames received increased.
- Link utilization is increased.

**MAC Layer Sat COM.**

- No. of frames sent increased.
- No. of frames received decreased – undesirable.
- No. of frames relayed increased.

**Network Layer IP**

- ip In received increased.
- ip In delivers increased.
- ip out request increased.
- ip In delivers TTL sum increased.

**Network Layer Strict Prior**

- No. of packets queued increased–Undesirable.
- No. of packets de-queued increased.

**Network Layer FIFO**

- Total packets queued increased –undesirable.
- Total packets de-queued increased.
- Peak queue size decreased.

**Transport Layer UDP**

- No. of packets from application layer increased.
- No. of packets to application layer increased.

**Application Layer Bellman Ford**

- No. of triggered updates increased.
- No.of update packets received increased.

## 5. CONCLUSION

Among the various parameters analyzed and compared with and without wireless Sensor nodes, 11 parameters have exhibited desirable improvement in performance, 4 parameters have exhibited undesirable change, 13 parameters have exhibited change which is not significant in performance evaluation. Hence an improvement by 73.33 % of Q.o.S is observed. In the case of WSN techniques for handoff it can be observed that undisturbed handoff and hence seamless roaming can be achieved almost every where. By employing these techniques any time connectivity may be dreamt off in the coming years. Satellite based location aware searching and therefore provision of services to the MS is also a feasible solution as proposed here.

## REFERENCES

[1] Wikipedia encyclopedia at http://en.wikipedia.org/wiki/Wireless_Sensor_Networks.

[2] Cross bow technologies ,USA Mote View User manual at http://www.xbow.com/Products/Product pdf_files/Wireless_pdf/MoteWorks_OEM_Edition.pdf

[3] S. Limm, G. Cao, C. R. Das, "*A differential bandwidth reservation policy for multimedia wireless networks*", Parallel Processing Workshops, International Conference on, 2001, pp. 447-452.

[4] D.A.Levine, I.F. Akyildiz, M. Naghshineh, "A resource, estimation and call admission algorithm for wireless multimedia networks using the shadow cluster concept". IEEE/ACM Transactions Networking **5** (1997), pp.1–12.

[5] Product tour and user's manual of Qualnet 4.5 by Scalable technologies, USA.

## BIOGRAPHY

N.S.V.Shet is presently working in the Dept of E & C ,NITK, Surathkal as faculty of Engg. He is a M.E from IIT roorkee. He is pursuing is PhD at NITK, Surathkal .His research interests are in the area of Wireless handoff and sensor networks. He has guided several Btech and Mtech projects. He was a visiting researcher at University Manchester Institute of science and technology, UK and Kagoshima University, Japan.

K.Chandra Sekaran is Professor in the Dept. of Computer Engg at NITK Surathkal. He has 22 years of teaching and academic research in the areas of networks, distributed computing, and over 95 publications in reputed journals / conferences.Incidently he is guiding N.S.V.Shet.

Prof. K.C. Shet has 30 years of teaching experience. His interests are in Research, Design, Development and Consulting (including Counseling) in the area of Computers, Electronics, Software Engineering, and Information / Network Security.